\documentclass[twoside]{LCWS11}
\usepackage[latin1]{inputenc}
\usepackage[dvips]{graphicx,epsfig,color}
\usepackage{wrapfig,rotating}
\usepackage{amssymb,amsmath,array}
\usepackage{cite}
\input paperdef
\begin{document}



\title{
The Higgs sector of the NMFV MSSM at the ILC} 

\author{M.~Arana-Catania$^{1}$, S.~Heinemeyer$^2$, M.J.~Herrero$^1$ and S.~Pe\~naranda$^{3,4}$
\thanks{Preprint number: IFT-UAM/CSIC-12-10}
\thanks{Talk given by M.~Arana-Catania at LCWS11}
\vspace{.3cm}\\
1- Departamento de F\'isica Te\'orica and Instituto de F\'isica Te\'orica,
IFT-UAM/CSIC\\
Universidad Aut\'onoma de Madrid, Cantoblanco, Madrid - Spain\\
\vspace{-.3cm}
\\
2- Instituto de F\'isica de Cantabria (CSIC-UC), Santander - Spain\\
\vspace{-.3cm}
\\
3- Departamento de F\'isica Te\'orica, Universidad de Zaragoza, Spain\\
\vspace{-.3cm}
\\
4- Departament de F\'isica Fonamental, Universitat de Barcelona, Spain\\
}

\maketitle

\begin{abstract}
We calculate the one-loop corrections to the Higgs boson masses within the context of the MSSM with Non-Minimal Flavor Violation in the squark sector. We take into account all the relevant restrictions from  \bsg, \bmm\ and \dmbs. We find sizable corrections to the lightest Higgs boson mass that are considerably larger than the expected ILC precision for acceptable values of the mixing parameters  $\deXYij$. We find $\de^{LR}_{ct}$ and $\de^{RL}_{ct}$ specially relevant, mainly at low $\tb$.
\end{abstract}

\section{Introduction}

We review the one-loop corrections to the Higgs boson masses in the MSSM
with Non-Minimal Flavor Violation (NMFV) \cite{AranaCatania:2011ak}. 
The flavor violation is generated from the hypothesis of general
flavor mixing in the squark mass matrices, parameterized
by a complete set of $\deXYij$ ($X,Y=L,R$; $i,j=t,c,u$ or $b,s,d$).
The corrections to the Higgs masses are calculated in terms of these 
$\deXYij$ taking into account all relevant restrictions from $B$-physics
data. In particular the present constraints from \bsg, \bmm\ and \dmbs are demanded to be fulfilled and our predictions are also compared within NMFV scenarios with the SM predictions. For completeness, we include below the present experimental data\cite{bexp}, and the predictions within the SM\cite{btheo}: 

\vspace{0.2cm}
\noindent \begin{equation}
\bsg_{\rm exp}=(3.55 \pm 0.26)\times10^{-4}\quad ;\quad \bsg_{\rm SM}= (3.15 \pm 0.23)\times10^{-4}
\label{bsgamma-exp}
\end{equation}

\noindent \begin{equation}
\bmm_{\rm exp} < 1.1 \times 10^{-8}\,\,\,\, (95\% ~{\rm CL}) ; \bmm_{\rm SM}= (3.6\pm 0.4)\times 10^{-9}
\label{bsmumu-SM}
\end{equation}

\noindent \begin{equation}
{\dmbs}_{\rm exp} = (117.0 \pm 0.8) \times 10^{-10} \mev~ \quad ;\quad {\dmbs}_{\rm SM} = (117.1^{+17.2}_{-16.4}) \times 10^{-10} \mev~.
\label{deltams-SM}
\end{equation} 

Here we focus on the analysis of the Higgs mass corrections that are originated from the flavor mixing between the second and third generations which is the relevant one in $B$ physics and devote special attention to the $LR/RL$ sector. These kind of mixing effects are expected
to be sizable, since they enter the off-diagonal $A$~parameters, which
appear directly in the coupling of the Higgs bosons to scalar quarks.

In the following we briefly review the main relevant aspects of the calculation and present the numerical results focusing on the light Higgs boson. For further details we address the reader to the full version
of our work \cite{AranaCatania:2011ak}, where also an extensive list with references to related works can be found.

\section{SUSY scenarios with Non-Minimal Flavor Violation}
\label{sec:nmfv}

The usual procedure to introduce general flavor mixing in the squark sector is to include the non-diagonality in flavor space in the Super-CKM basis. These squark flavor mixings are usually described in terms of a set of dimensionless parameters $\deXYij$ ($X,Y=L,R$; $i,j=t,c,u$ or $b,s,d$), introduced in the SUSY-breaking matrices (after RGE running) at low energy as follows,

\noindent \begin{equation}  
m^2_{\tilde U_L}= \left(\begin{array}{ccc}
 m^2_{\tilde U_{L11}} & 0 & 0\\
0 & m^2_{\tilde U_{L22}}  & \delta_{23}^{LL} m_{\tilde U_{L22}}m_{\tilde U_{L33}}\\
0 & \delta_{23}^{LL} m_{\tilde U_{L22}}m_{\tilde U_{L33}}& m^2_{\tilde U_{L33}} \end{array}\right);m^2_{\tilde D_L} =\VCKM^{\dagger}m^2_{\tilde U_L}\VCKM\end{equation}

\noindent \begin{equation}
v_2 {\cal A}^u  =\left(\begin{array}{ccc}
0 & 0 & 0\\
0 & 0 & \delta_{ct}^{LR} m_{\tilde U_{L22}}m_{\tilde U_{R33}}\\
0 & \delta_{ct}^{RL} m_{\tilde U_{R22}}m_{\tilde U_{L33}} & m_{t}A_{t}\end{array}\right)\end{equation}

\noindent \begin{equation}  
m^2_{\tilde U_R}= \left(\begin{array}{ccc}
 m^2_{\tilde U_{R11}} & 0 & 0\\
0 & m^2_{\tilde U_{R22}}  & \delta_{ct}^{RR} m_{\tilde U_{R22}}m_{\tilde U_{R33}}\\
0 & \delta_{ct}^{RR} m_{\tilde U_{R22}}m_{\tilde U_{R33}}& m^2_{\tilde U_{R33}} \end{array}\right)\end{equation}

and analogously for $v_1 {\cal A}^d$ and $m^2_{\tilde D_R}$, changing the up-type indexes to the down-type ones in $v_2 {\cal A}^u$ and $m^2_{\tilde U_R}$, correspondingly. The flavor diagonal entries in these matrices at low energy are found here as usual, namely  after RGE running  and assuming universality conditions for the soft parameters at the GUT scale (i.e. within constrained models).

In the present study we will restrict ourselves
  to the case where there is flavor mixing exclusively between the second and third squark generation. These mixings are known to produce the largest flavor violation effects in $B$ meson physics since their size are usually governed by the third generation quark masses. On the other hand, and in order to reduce further the number of independent parameters, we will focus in the following analysis on 
 constrained  SUSY scenarios, where the soft mass parameters fulfill universality hypothesis at the gauge unification (GUT) scale. Concretely, we will restrict ourselves here to the so-called Constrained MSSM (CMSSM) which is defined by $m_0, m_{1/2}, A_0, {\rm sign}(\mu), \tb$, where $A_0$ is the universal trilinear coupling, $m_0$ and $m_{1/2}$  are the universal scalar mass and
gaugino mass, respectively, at the GUT scale,   ${\rm sign}(\mu)$ is the sign of the $\mu$ parameter and $\tb =v_2/v_1$. 

For the following numerical estimates we will chose two particular points in the CMSSM that are examples of scenarios with moderate and very heavy sparticles masses, respectively. Firstly, we set the well-known benchmark point SPS2, with $m_0=1450~ {\rm GeV},~ m_{1/2}=300~ {\rm GeV},~ A_0=0,~ {\rm sign}(\mu)>0,~ \tb=10$. Secondly, we study a peculiar scenario, nowadays favored by LHC recent data, where the SUSY particles are rather heavy, at the TeV scale, but still the Higgs particle is light. We name this scenario as VHeavyS, defined by $m_0=m_{1/2}=-A_0=800~ {\rm GeV},~ {\rm sign}(\mu)>0,~ \tb=5$ and where $m_{h}^{\rm MSSM}=120 {\rm GeV}$. 
The corresponding analysis for other points in the CMSSM and other scenarios as the Non Universal Higgs Mass case can be found in \cite{AranaCatania:2011ak}.

\section{Results} 
\label{sec:Bphysics}

In this section we review our numerical results for the radiative corrections to the Higgs boson
 mass $m_{h}$ from flavor mixing within NMFV-SUSY scenarios. Since all one-loop corrections in the present NMFV scenario are common to the MSSM except for the corrections from squarks, which depend on the $\deXYij$ values,  we will focus just on the results of these corrections as a function of the flavor mixing parameters,  and present the differences with respect to the predictions within the MSSM. Correspondingly, we define: 
\begin{equation}
 \Dmh (\deXYij) \equiv 
 m_{h}^{\rm NMFV}(\deXYij) - m_{h}^{\rm MSSM} 
\end{equation}
where $m_{h}^{\rm NMFV}(\deXYij)$ and $m_{h}^{\rm MSSM}$ have been calculated at the
one-loop level.
 It should be noted that  $m_{h}^{\rm NMFV}(\deXYij=0) = m_{h}^{\rm MSSM}$ and,
 therefore, by construction,  $\Dmh(\deXYij=0) = 0$, and  $\Dmh$ gives the
size of the one-loop NMFV contributions to $m_{h}$. The numerical calculation of $m_{h}^{\rm NMFV}(\deXYij)$ and $m_{h}^{\rm MSSM}$ has been done 
with \fh~\cite{feynhiggs}. The numerical calculations of the rates for the $B$ observables have been done with the FORTRAN subroutine BPHYSICS included in the SuFla code\cite{sufla}, which we have conveniently modified as to include all the relevant contributions within NMFV scenarios (for more details see  \cite{AranaCatania:2011ak}).

\begin{figure}[ht!]
   \begin{center} 
     \begin{tabular}{cc} \hspace*{-8mm}
        \psfig{file=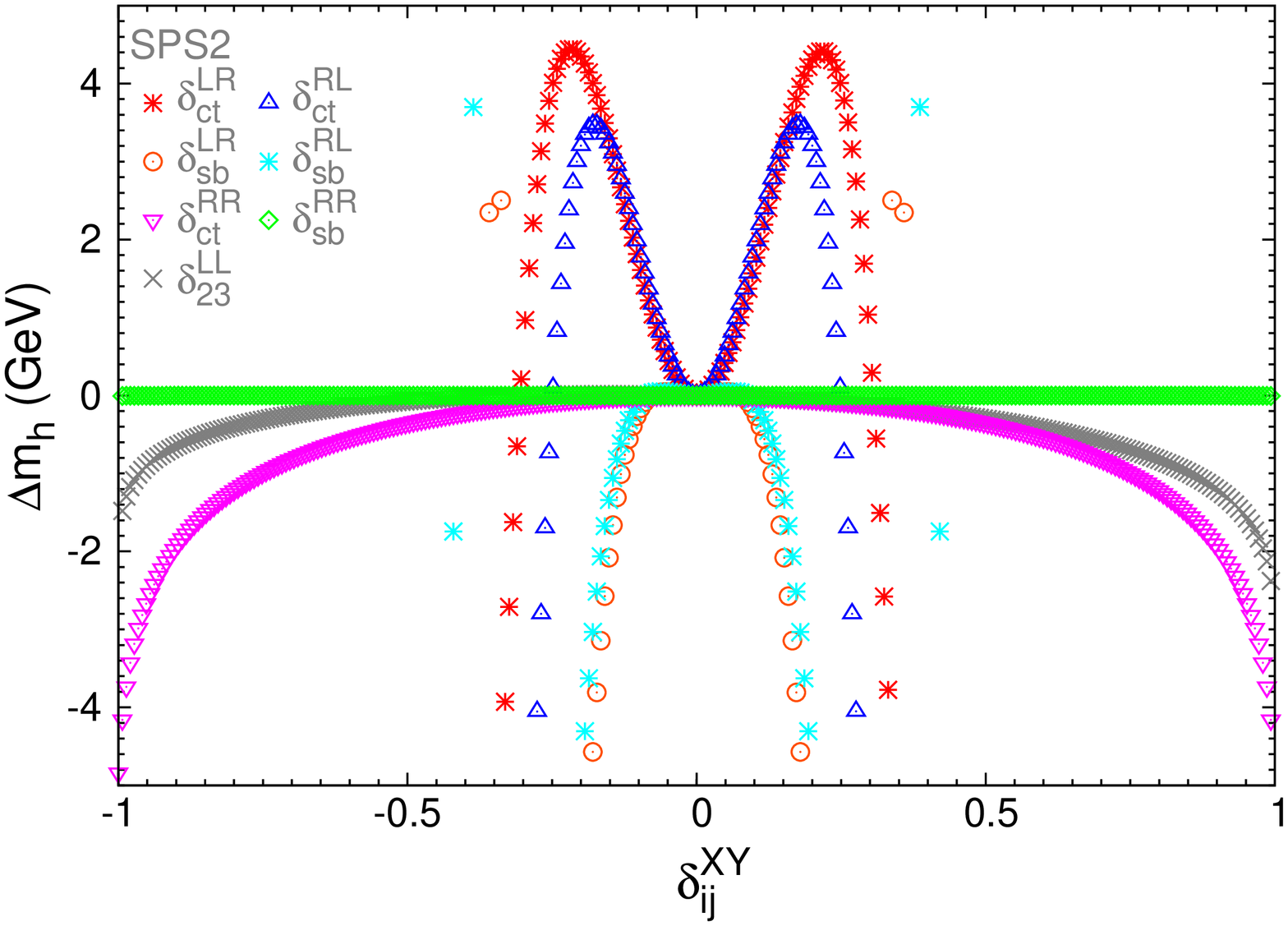,width=50mm,angle=270,clip=} 
        &
        \psfig{file=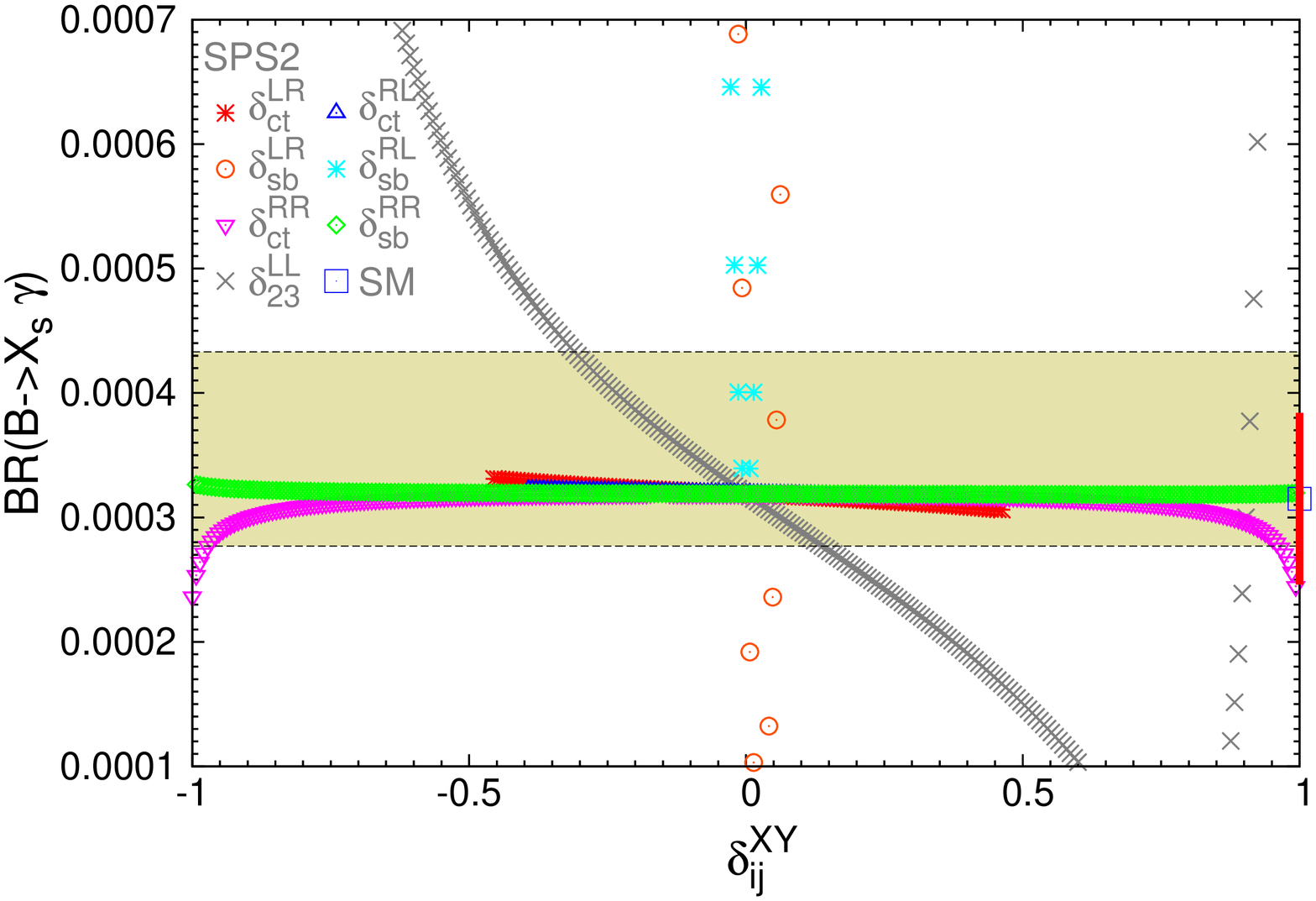,width=50mm,angle=270,clip=}\\         
        \psfig{file=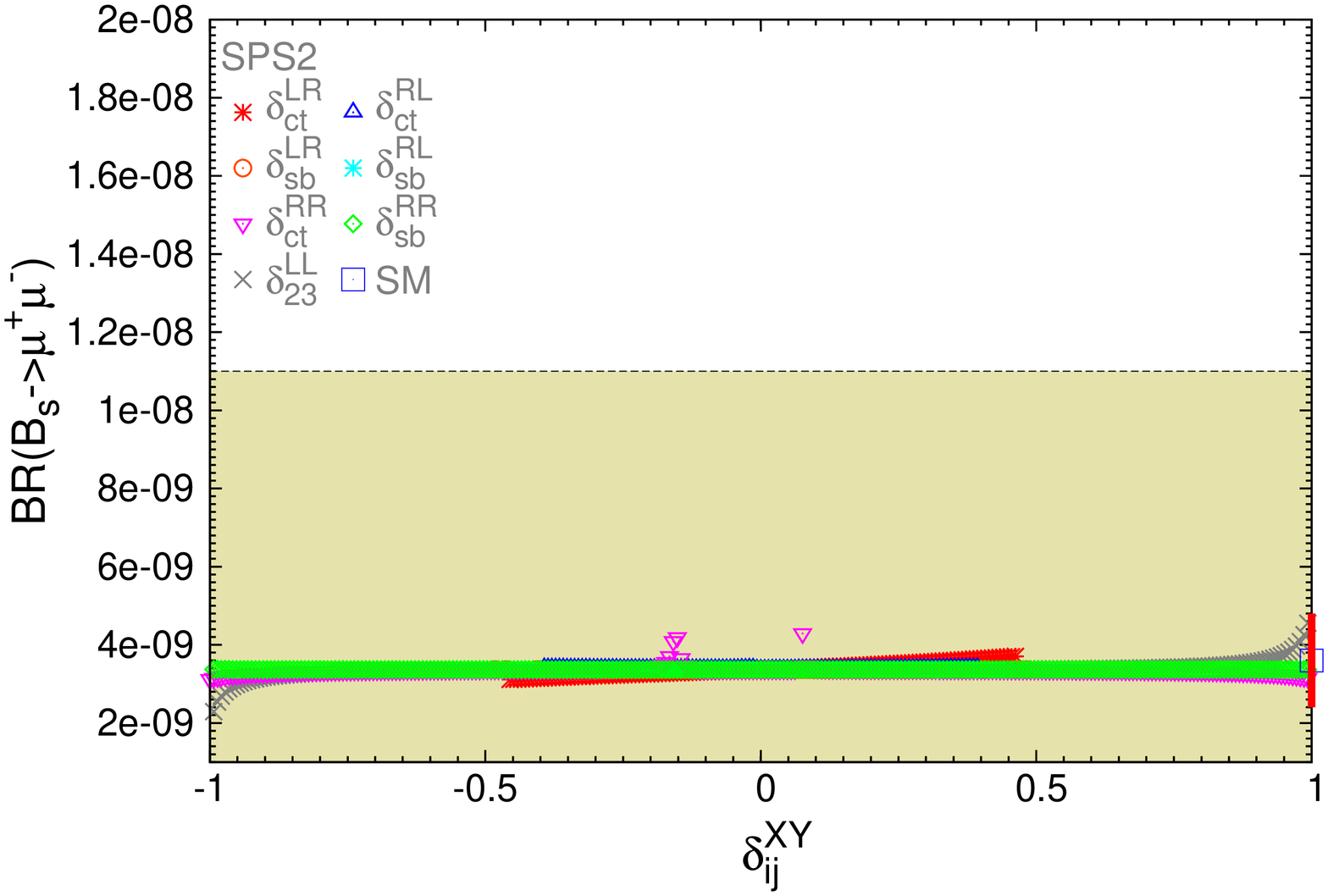,width=50mm,angle=270,clip=} 
        &
        \psfig{file=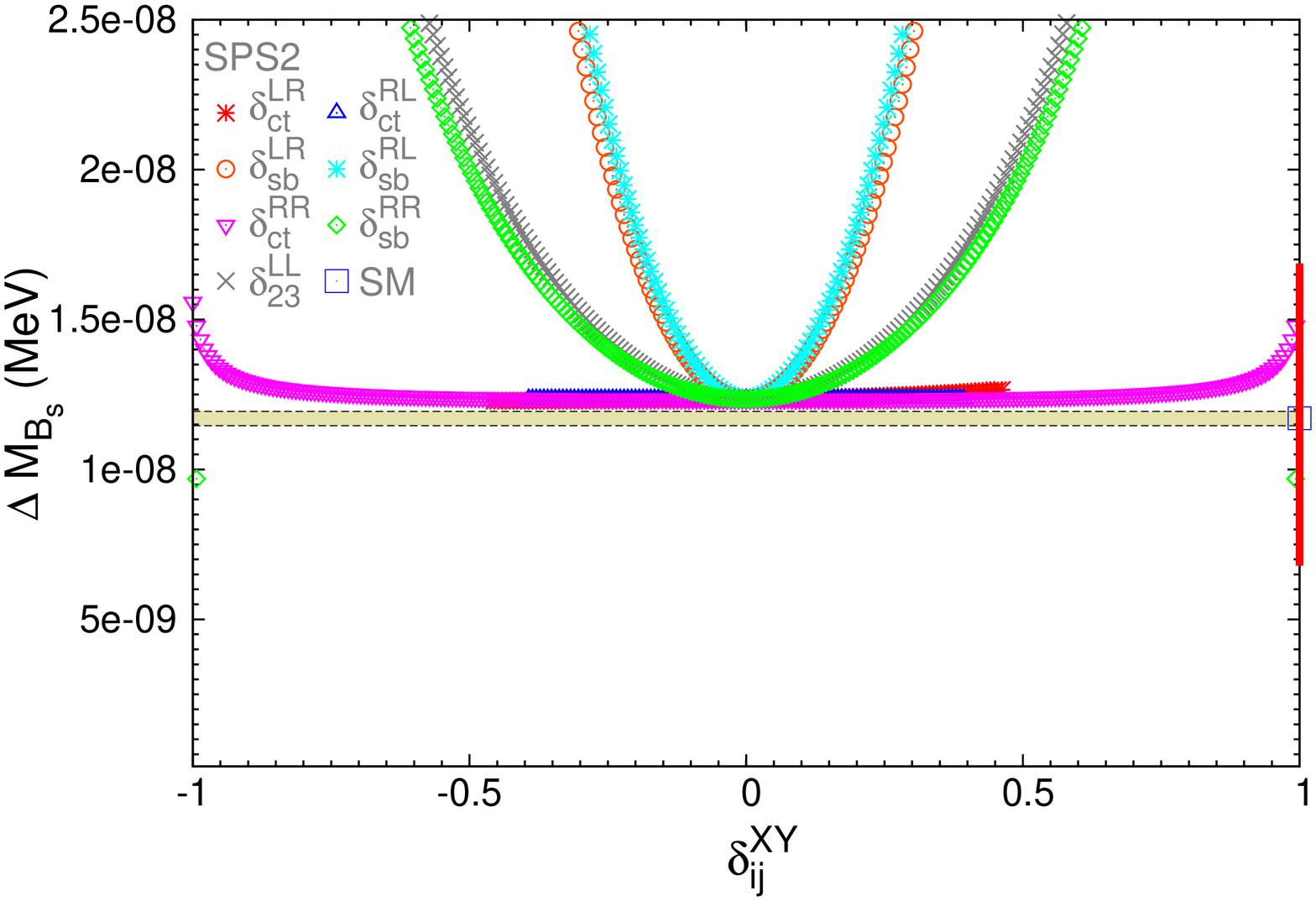,width=50mm,angle=270,clip=}         
       \end{tabular}
\caption{Sensitivity to the NMFV deltas for the SPS2 point for different observables:  $\Delta m_{h}$  (left upper panel), \bsg\  (right upper panel),  \bmm\  (left bottom panel), \dmbs  (right bottom panel). The experimental allowed areas in the plot for $B$~physics observables ($3\sigma_{\rm exp}$ for \bsg\ and \dmbs, $95\% ~{\rm CL}$ bound for \bmm\ ) are the horizontal colored bands. The SM prediction for the $B$~physics observables and the theory uncertainty (red bar) is displayed on the right axis, correspondingly.}  
     \label{fig:sps2fig} 
   \end{center}
 \end{figure}

In Fig.\ref{fig:sps2fig} we show the numerical results for $\Dmh$ as a function of the various $\deXYij$.  We have also included our predictions for  \bsg, \bmm\ and \dmbs  and their corresponding experimental allowed areas.

In order to conclude on the allowed delta intervals by $B$~physics data, we have assumed that our
predictions of the $B$ observables within SUSY-NMFV scenarios have a somewhat larger theoretical error than the SM prediction. Then, by adding linearly the experimental uncertainty (that we take as $3\sigma_{\rm exp}$) 
and the theoretical uncertainty, a given $\deXYij$ value is considered by us to be allowed
by data if the total predicted ratios lie in the following intervals: 
 \noindent \begin{align}
\label{linearerr}
2.08\times 10^{-4} < \bsg < 5.02\times 10^{-4},\\ \bmm < 1.22 \times 10^{-8},\\ 63\times 10^{-10} < \dmbs {\rm (MeV)} < 168.6\times 10^{-10}.
 \end{align} 

In table \ref{tableintervals} we summarize the total allowed delta intervals by ${\rm BR}(B\rightarrow X_s\gamma)$, ${\rm BR}(B_s\rightarrow\mu^+\mu^-)$ and $\dmbs$ for SPS2. Notice that for $\delta^{RR}_{sb}$ there are two very narrow allowed region close to $\pm 1$, which indeed for SPS2 reduce just to the two single allowed values $\pm 0.99$. 

\begin{wraptable}{l}{0.5\columnwidth}
\begin{center}
\begin{tabular}{|c|c|} \hline
 SPS2 & Total allowed intervals \\ \hline
$\delta^{LL}_{23}$  &  
\begin{tabular}{c} 
 (-0.37:0.34) \end{tabular} \\
 \hline
$\delta^{LR}_{ct}$      
& \begin{tabular}{c} 
 (-0.46:0.46) \end{tabular}   \\ \hline
$\delta^{LR}_{sb}$      & 
\begin{tabular}{c} 
 (-0.0069:0) (0.048:0.055) \end{tabular}  \\ \hline
$\delta^{RL}_{ct}$      & 
\begin{tabular}{c}
 (-0.39:0.39) \end{tabular} 
  \\ \hline
$\delta^{RL}_{sb}$      &
\begin{tabular}{c}  (-0.014:0.014) \end{tabular} 
  \\ \hline
$\delta^{RR}_{ct}$    & \begin{tabular}{c} 
 (-1.0:0.99) \end{tabular}    \\ \hline
$\delta^{RR}_{sb}$      &
\begin{tabular}{c}  (-0.99) (-0.39:0.39) (0.99)
\end{tabular}    \\ \hline
\end{tabular}  
\end{center}
\caption{Allowed delta intervals by ${\rm BR}(B\rightarrow X_s\gamma)$, ${\rm BR}(B_s\rightarrow\mu^+\mu^-)$ and $\dmbs$ for SPS2. \label{tableintervals}}
\end{wraptable}

As we can see in Fig.\ref{fig:sps2fig}, the most restrictive observables are \bsg~  and $\Delta M_{B_s}$, leading to the total allowed delta intervals summarized in table \ref{tableintervals}. The main conclusion from this table is that the NMFV deltas in the top-sector can be sizeable $|\de_{ct}^{XY}|$, larger than ${\cal O}(0.1)$ and still compatible with $B$~physics data.  In particular $\delta^{RR}_{ct}$ is the less constrained parameter. The parameters on the bottom-sector are, in contrast, quite constrained. The most tightly constrained are clearly  $\delta^{LR}_{sb}$ and  $\delta^{RL}_{sb}$. Similar conclusions are found for other CMSSM points studied in \cite{AranaCatania:2011ak}.

Regarding the size of the mass corrections, $\Dmh$, we clearly see in this figure that they can be sizable for non-vanishing deltas in the allowed intervals by $B$~physics data. In particular for $\delta^{LR}_{ct}$ and $\delta^{RL}_{ct}$ they can be positive and up to about 4 GeV or negative and up to tens of GeV. $\delta^{RR}_{ct}$ yields negative corrections and up to about 4 GeV.

\begin{figure}[h!] 
\centering
\hspace*{-10mm} 
{\resizebox{13.5cm}{!} 
{\begin{tabular}{cc} 
\includegraphics[width=13.3cm]{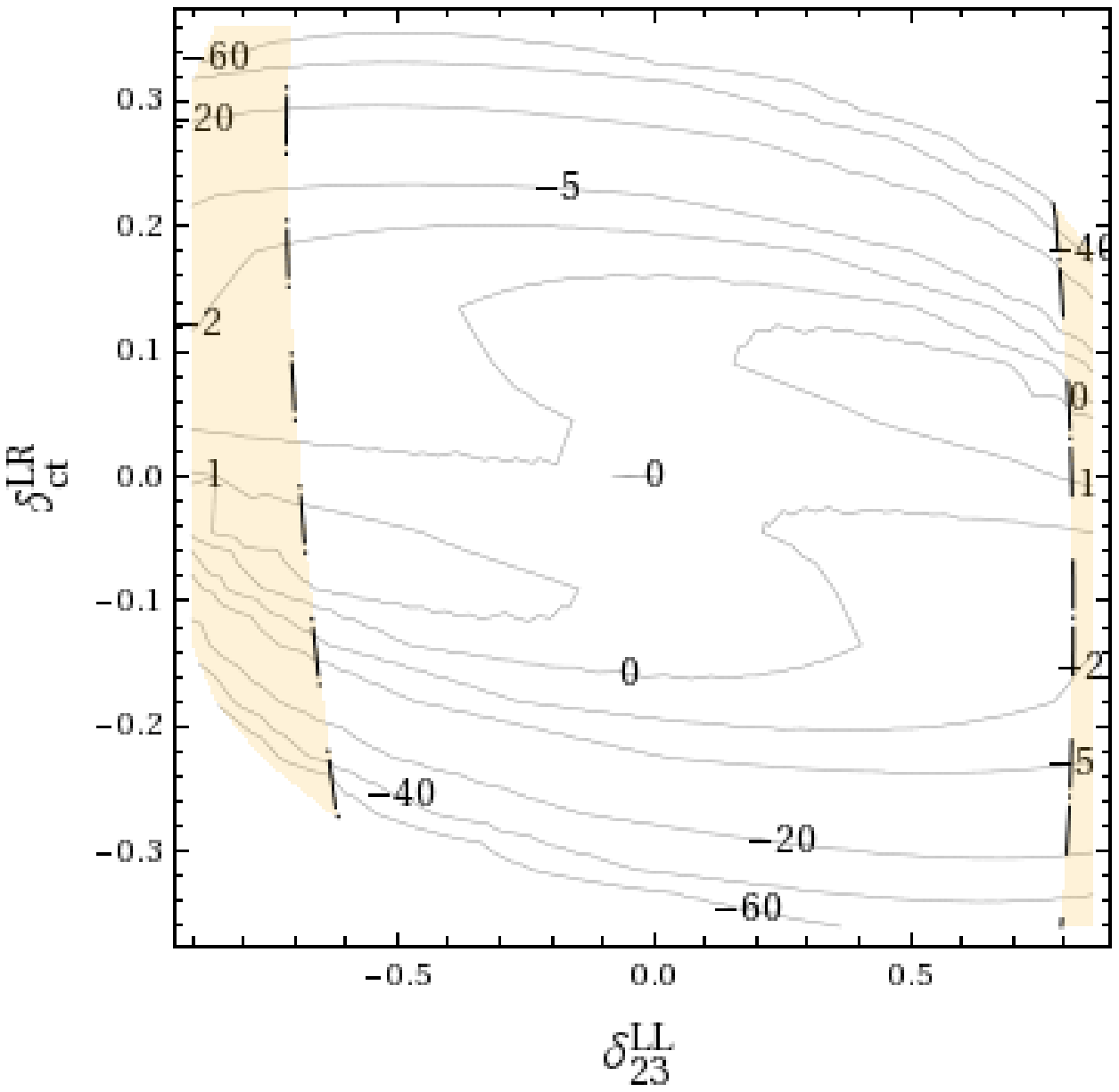}& 
\includegraphics[width=13.3cm]{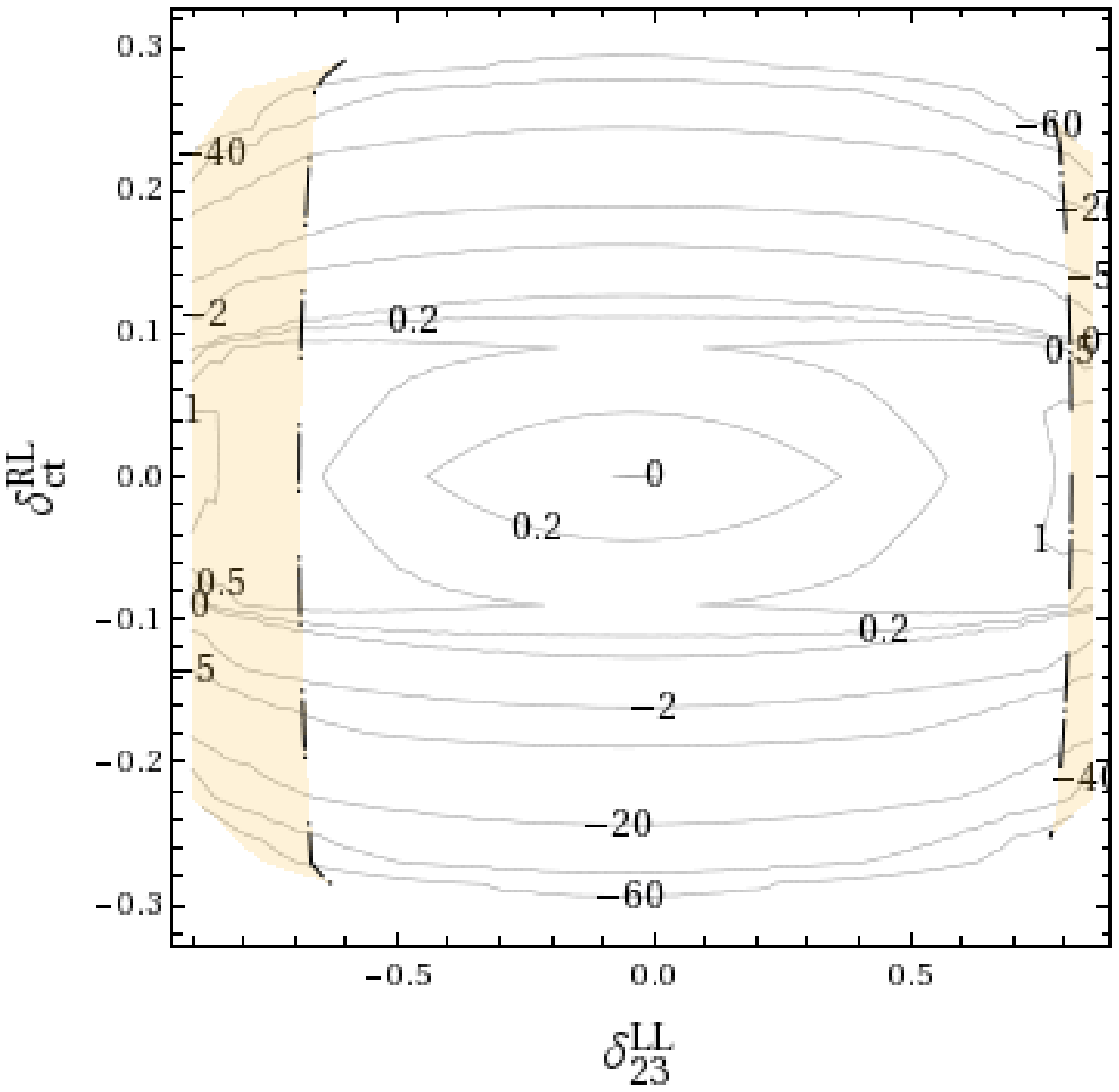}\\ 
\includegraphics[width=13.3cm]{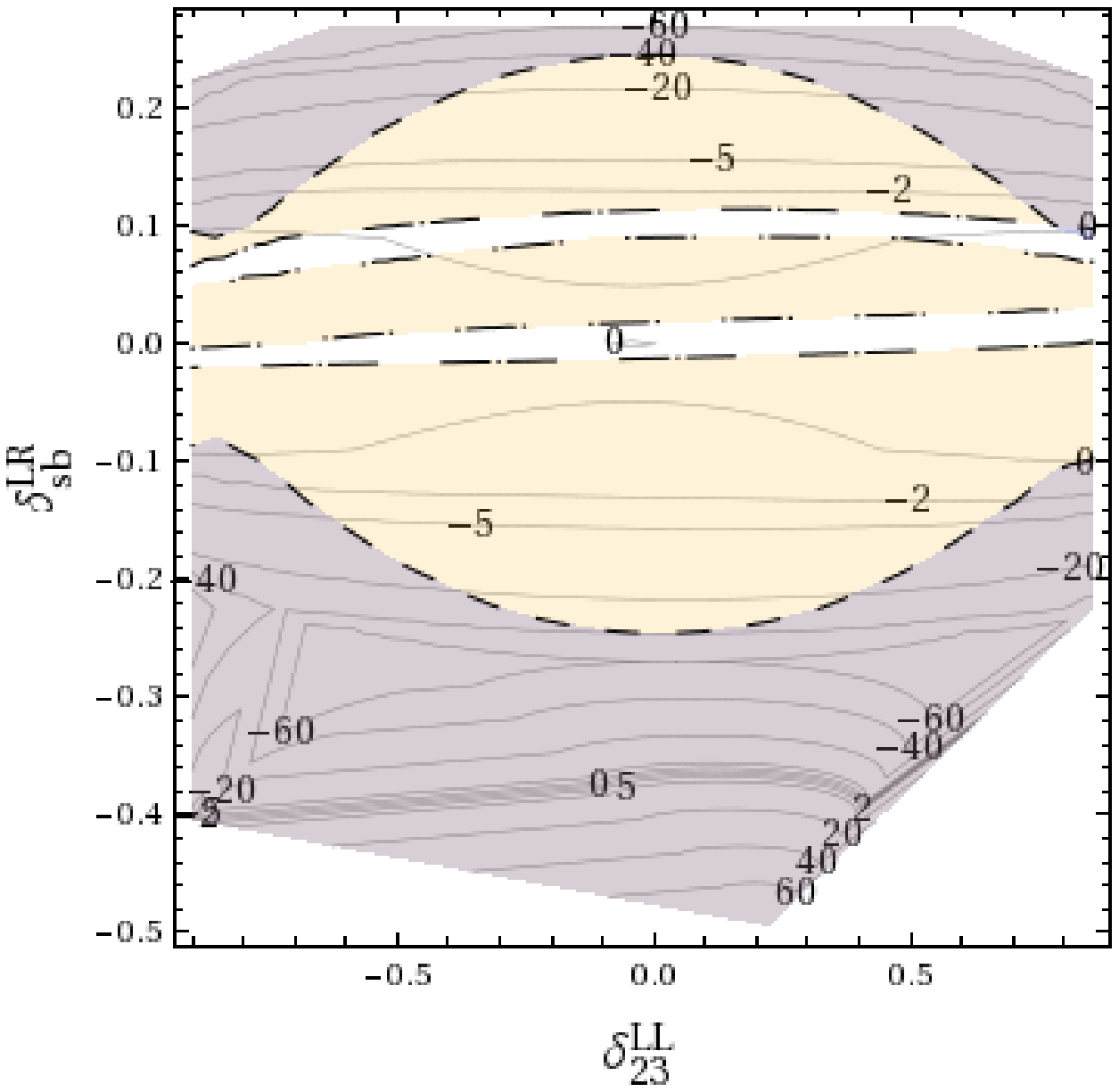}&
\includegraphics[width=13.3cm]{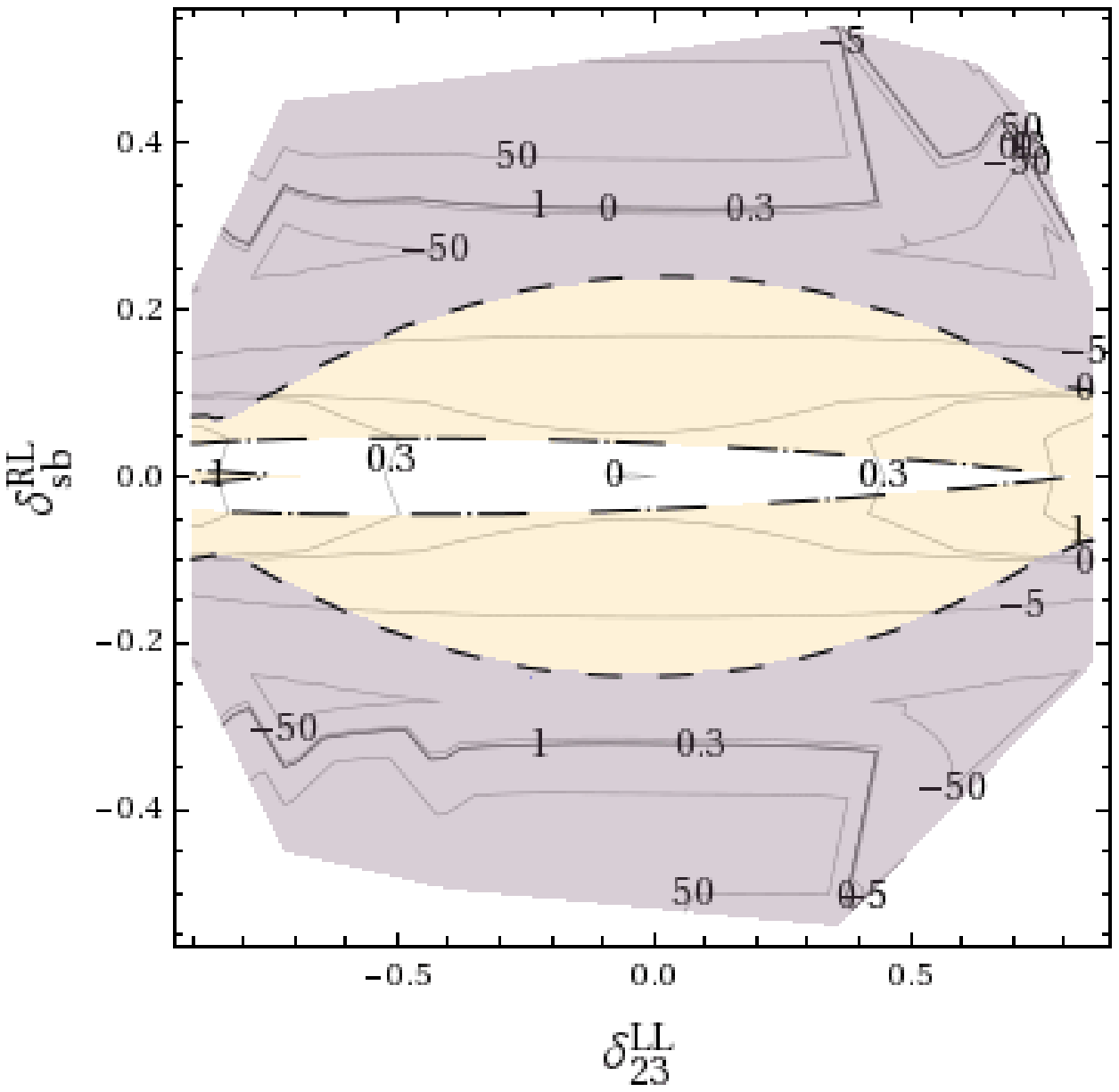}\\ 
\includegraphics[width=13.3cm]{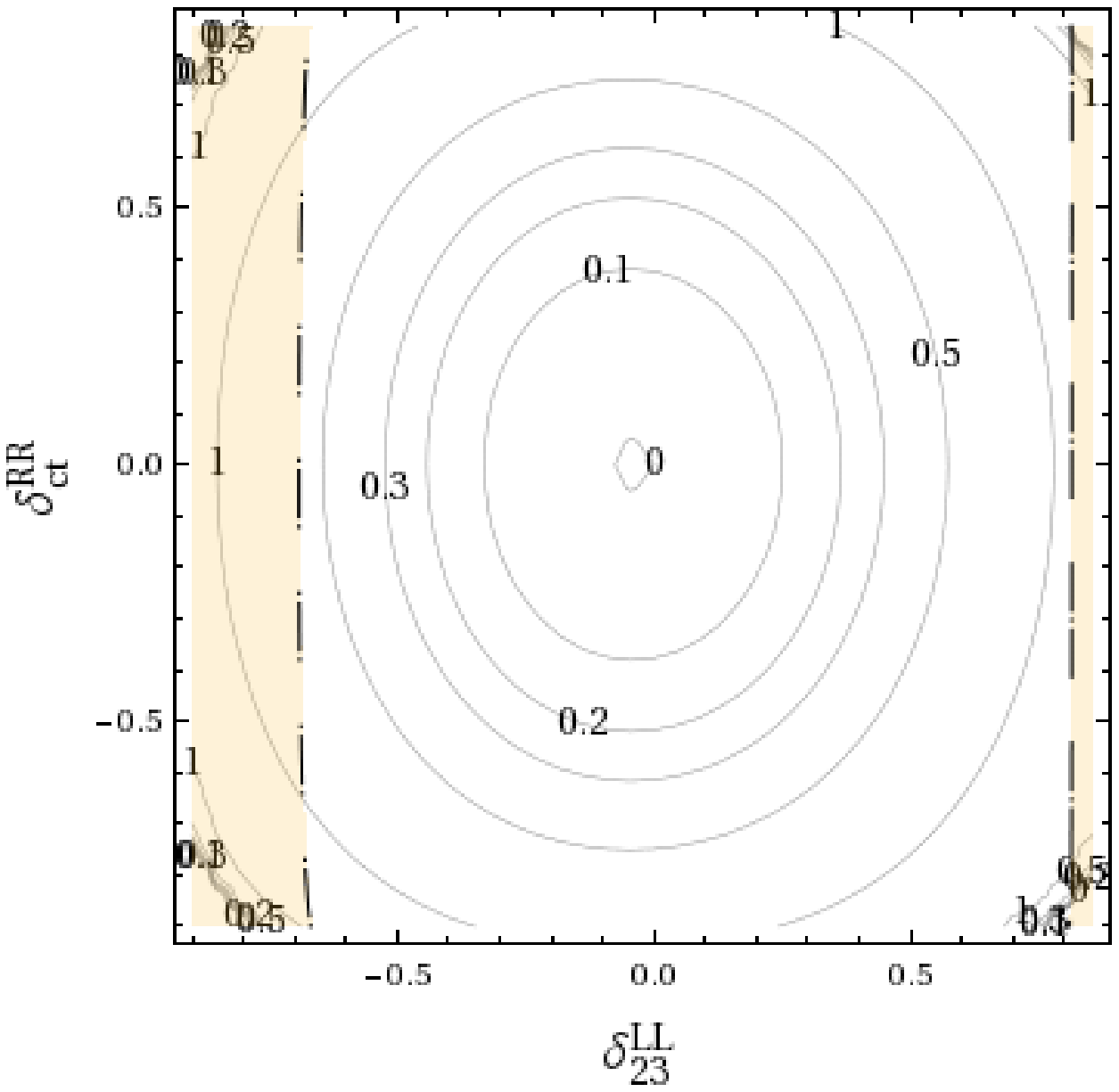}& 
\includegraphics[width=13.3cm]{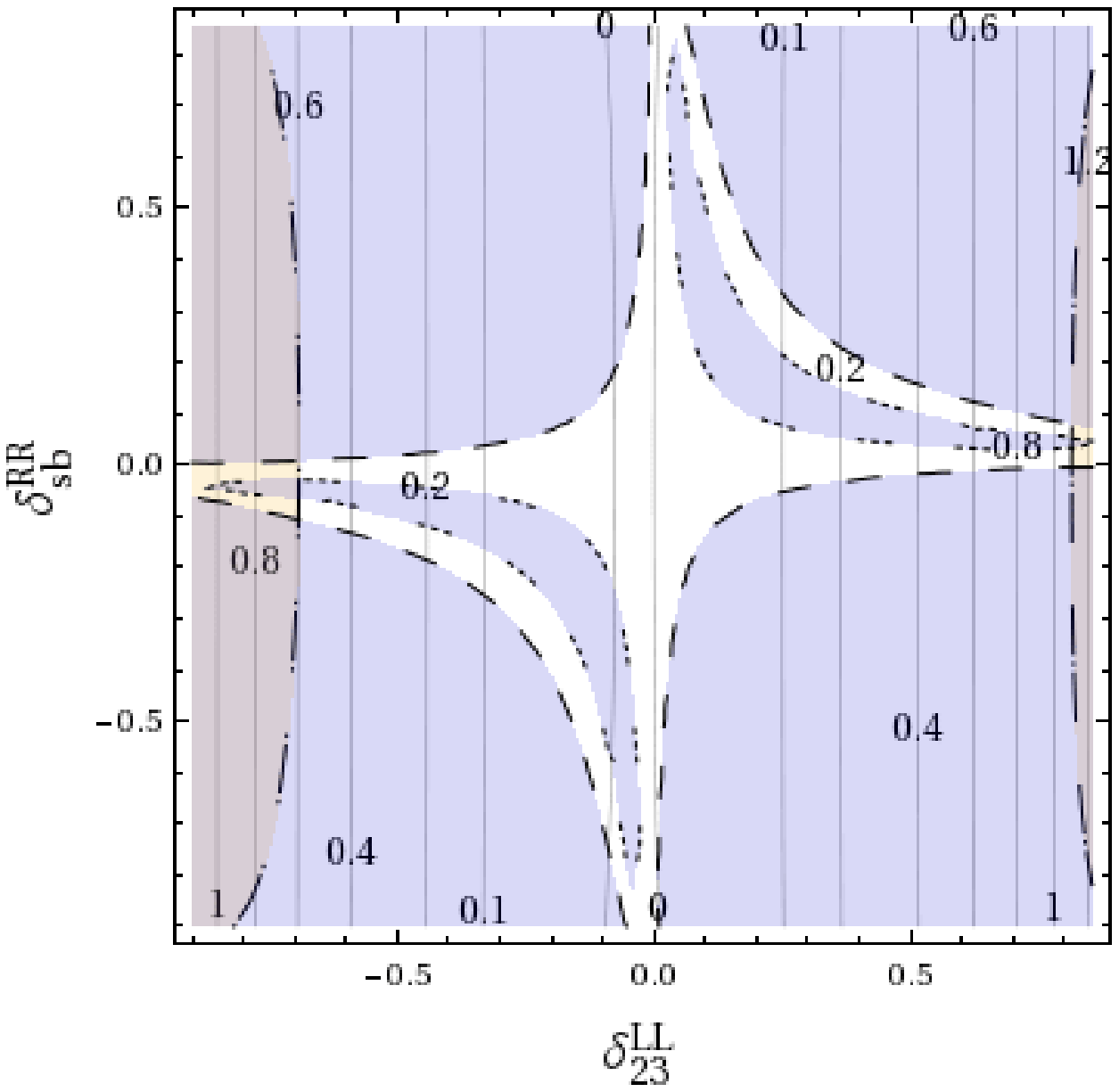}\\ 
\end{tabular}}}
\caption{$\Delta m_{h}$ (GeV) contour lines from our two deltas analysis for VHeavyS. The color code for the allowed/disallowed areas by $B$~physics data is given in fig.\ref{fig:colleg}} 
\label{figVHeavyS}
\end{figure}

\begin{figure}[h]  
\centering 
{\resizebox{10cm}{!}  
{\begin{tabular}{cccc}  
\includegraphics[width=13.3cm]{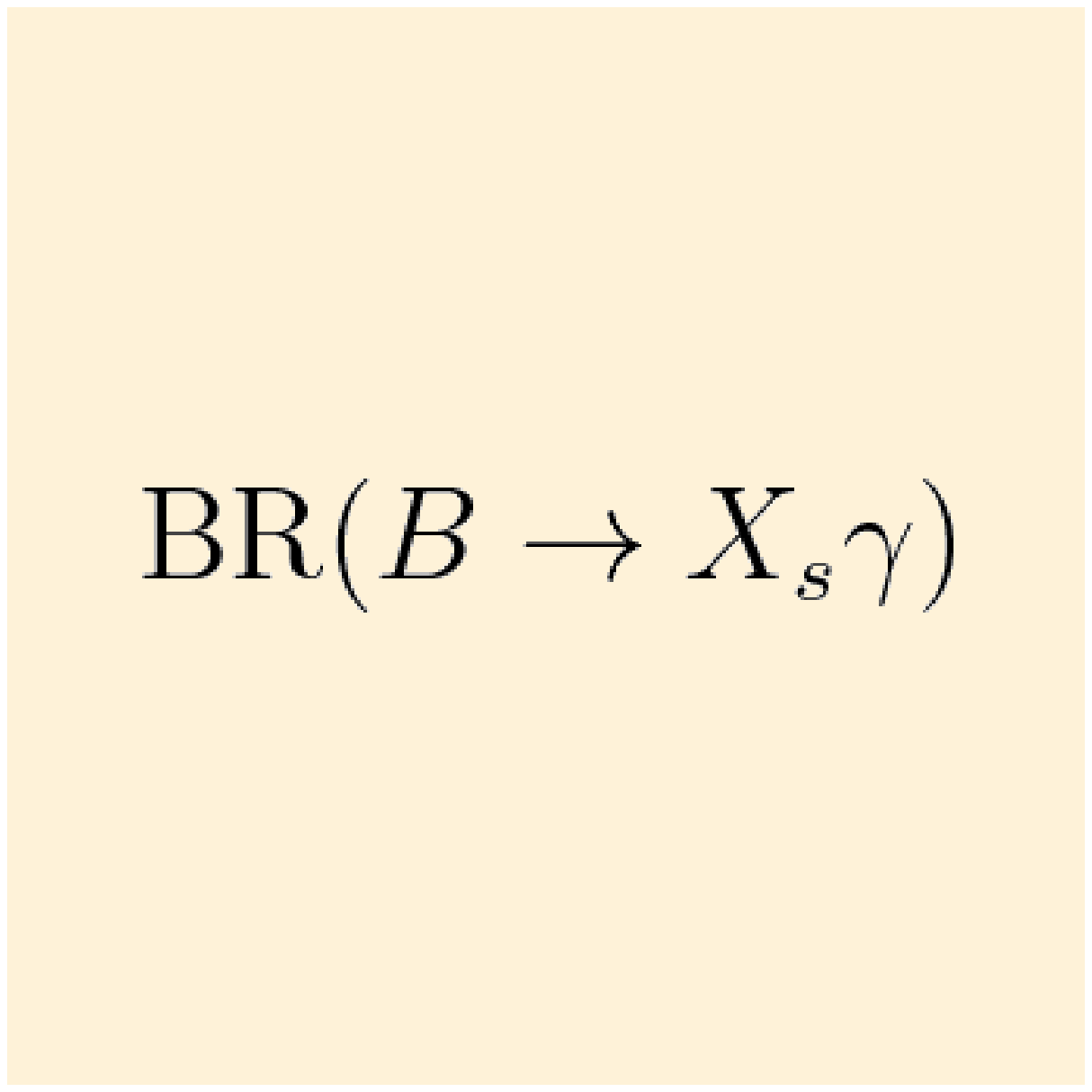}&  
\includegraphics[width=13.3cm]{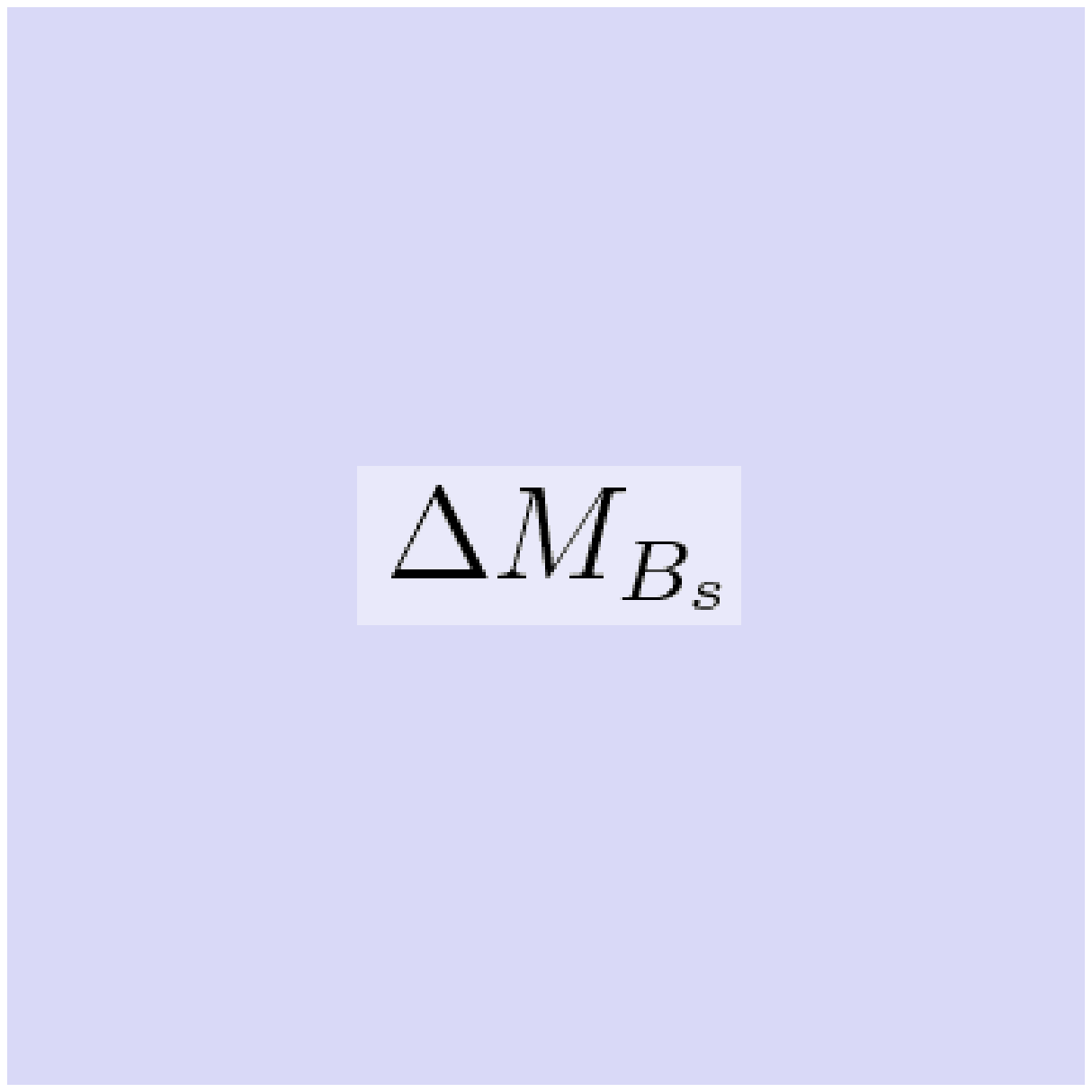}&  
\includegraphics[width=13.3cm]{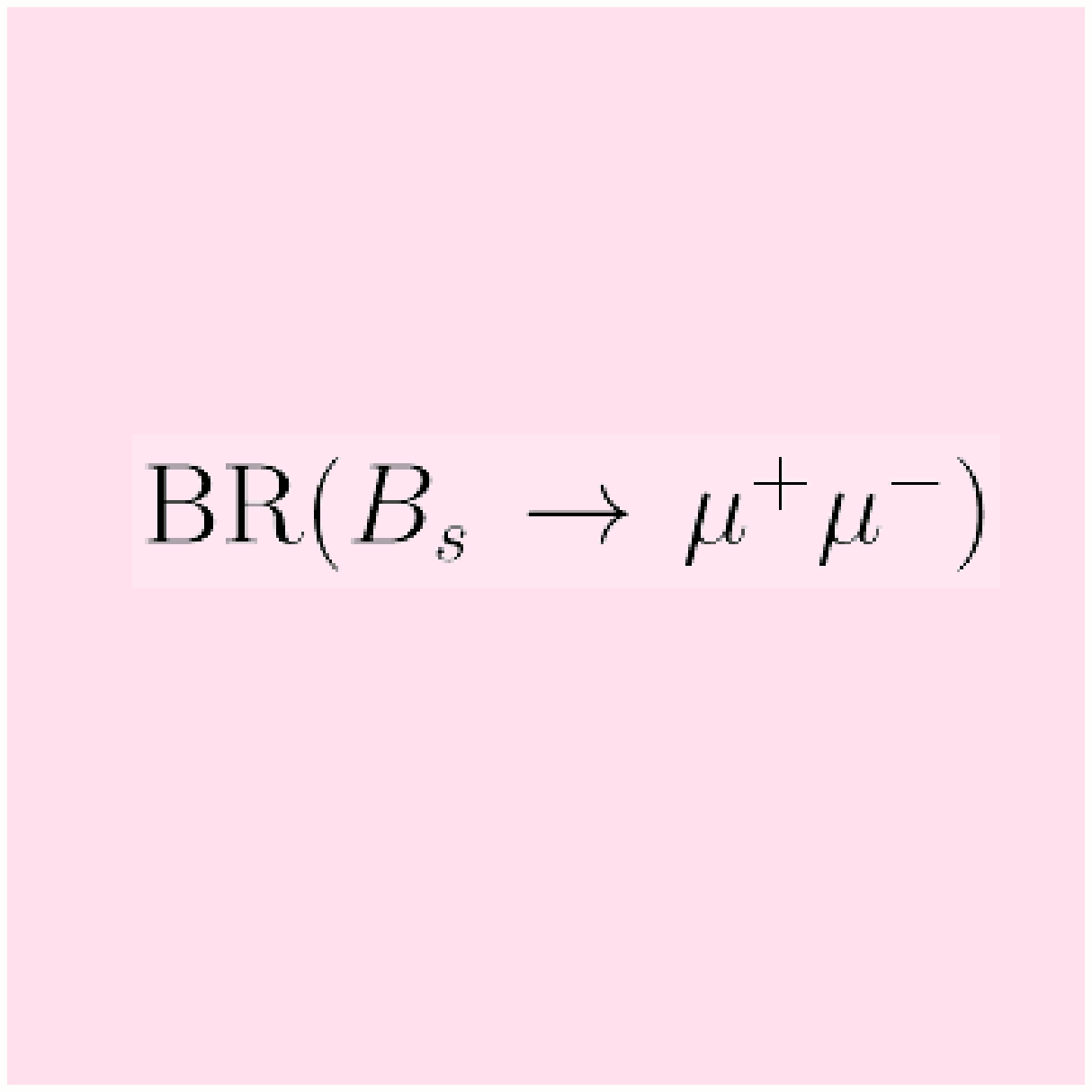}& 
\includegraphics[width=13.3cm]{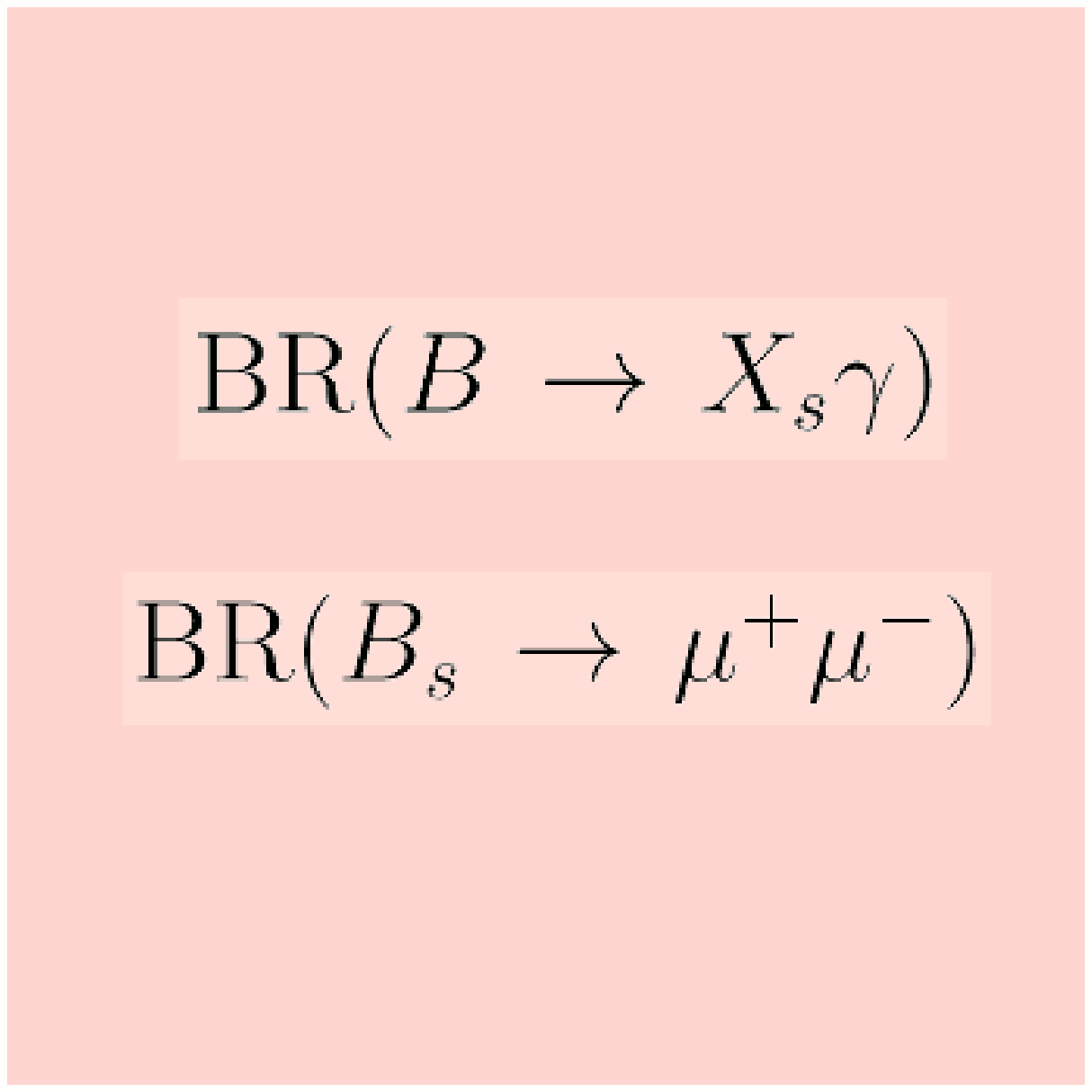}\\
\includegraphics[width=13.3cm]{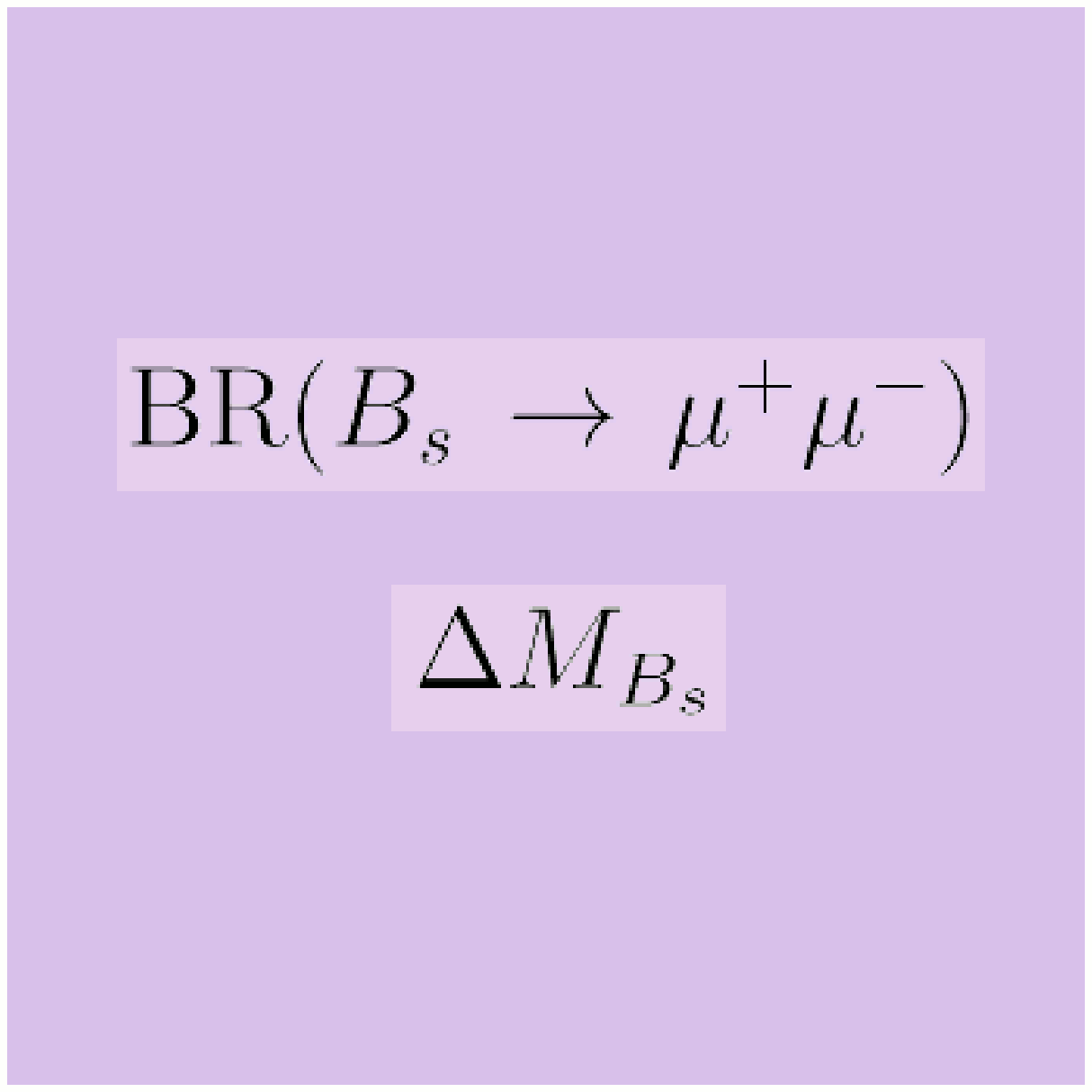}&  
\includegraphics[width=13.3cm]{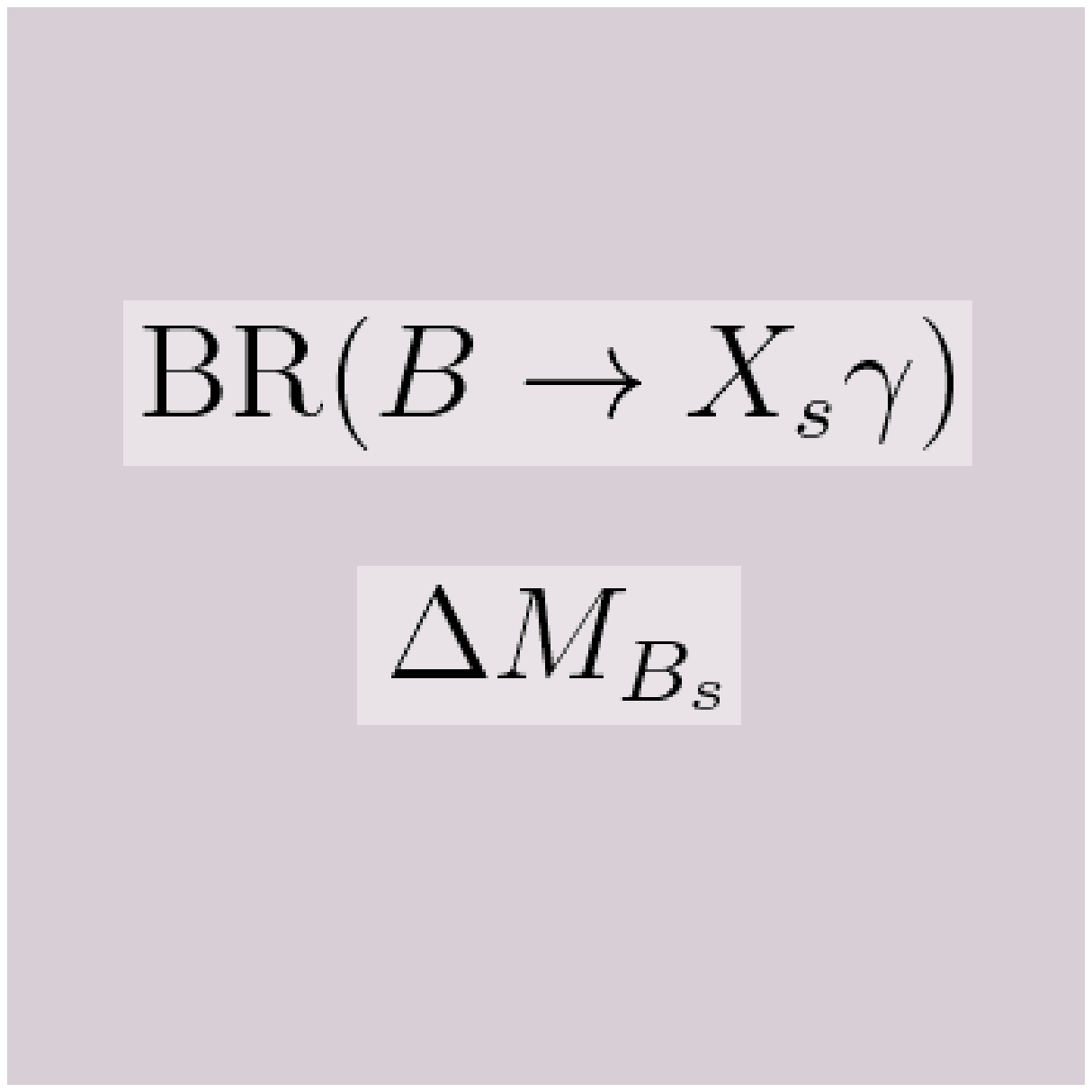}& 
\includegraphics[width=13.3cm]{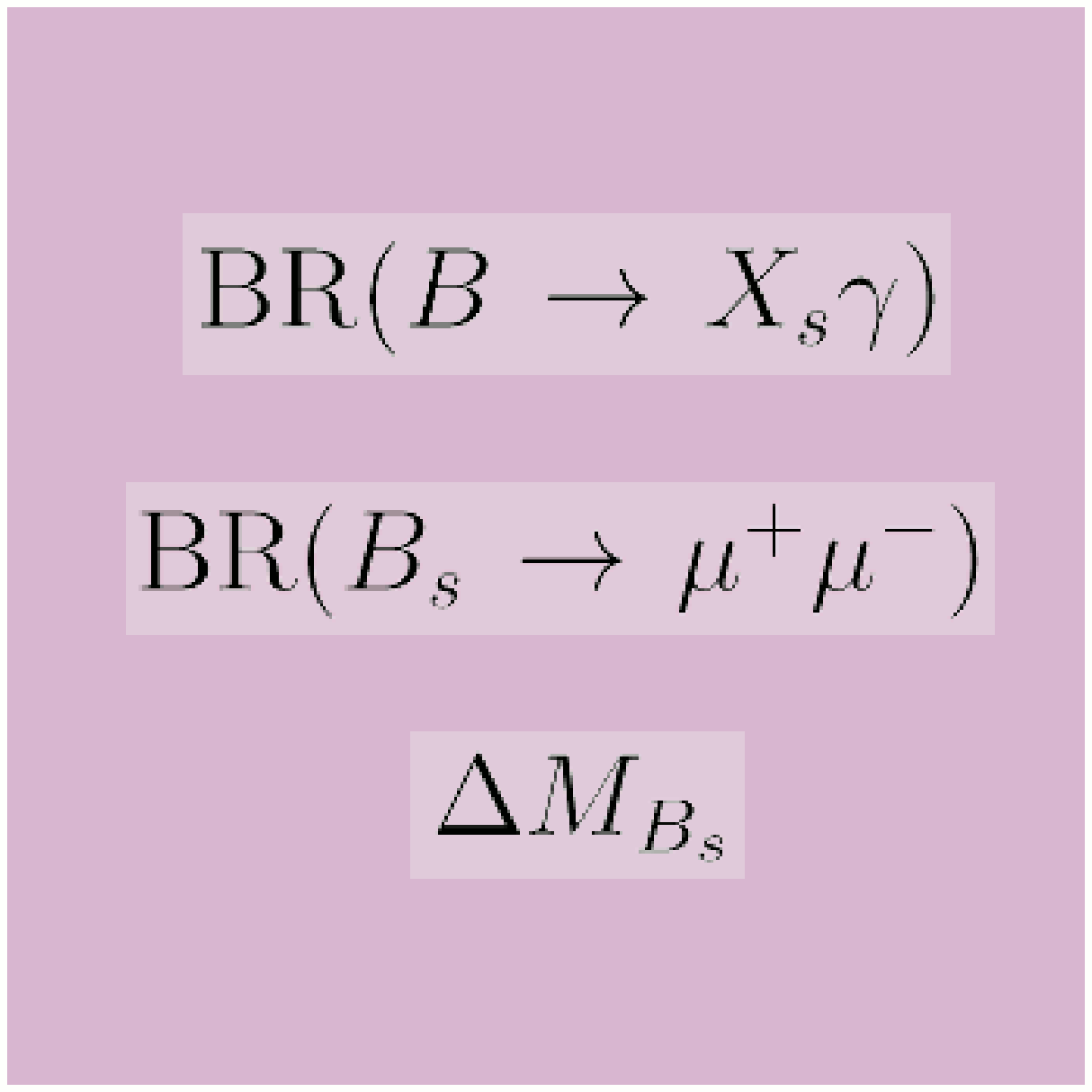}&
\includegraphics[width=13.3cm]{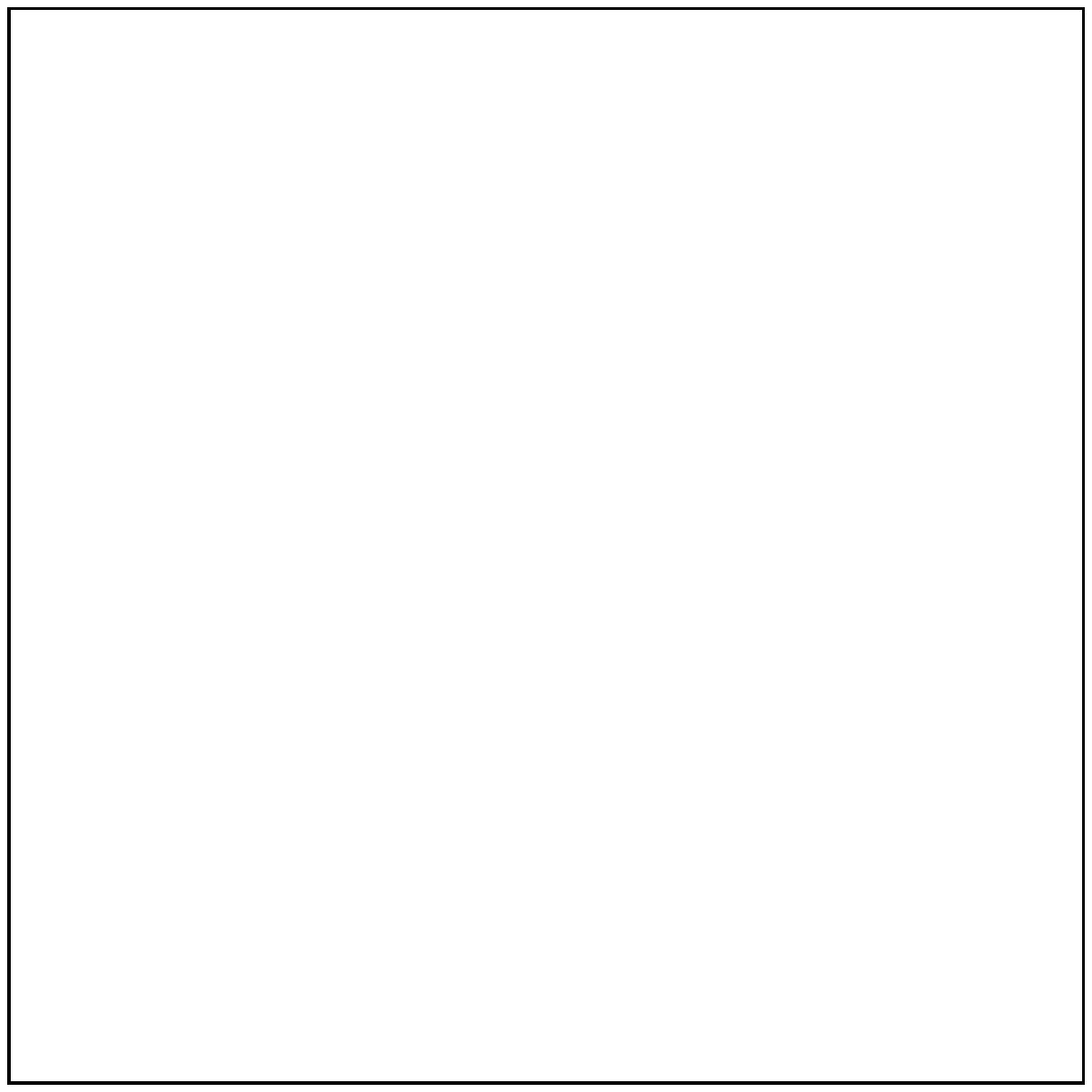}\\
\end{tabular}}} 
\caption{Legend of plots for Fig. \ref{figVHeavyS}. Each colored area represents the disallowed region by the specified observable/s inside each box. A white central area represents a region that is allowed by all $B$~physics data}  
\label{fig:colleg} 
\end{figure}

In order to explore further the size of the Higgs mass corrections, we have computed numerically the size of  $\Dmh$ as a function of two non-vanishing deltas and have looked for areas in these two dimensional plots that are allowed by $B$~physics data (see Fig.\ref{fig:colleg} for the color code of the allowed/disallowed areas). The results for VHeavyS are displayed in  Fig.\ref{figVHeavyS}.  Contour lines corresponding
to mass corrections $\Dmh$ above 60 GeV or below -60 GeV have not been represented.

\clearpage
\newpage
\clearpage

As we can see, the largest mass corrections $\Dmh$ found, being allowed by $B$~physics data occur in the $(\de^{LL}_{23},\de^{LR}_{ct} )$ and  $(\de^{LL}_{23},\de^{RL}_{ct} )$ planes. This applies also to the other points studied in \cite{AranaCatania:2011ak}.
They can be as large as $-60$ GeV for $\de^{LR}_{ct}$ or $\de^{RL}_{ct}$ close to $\pm 0.3$. Again these large corrections from the $LR$ and $RL$ parameters are due to the $A$-terms.
Generically, the plots  with largest
allowed regions and with largest Higgs mass corrections correspond to
scenarios with low $\tb = 5$ 
and heavy spectra like the VHeavyS point considered here.

There are also important corrections in the allowed areas of the two dimensional plots of   $(\de^{LL}_{23},\de^{RR}_{ct} )$ for some of the studied points, particularly for SPS5 (and to a lesser extent for SPS2). For the first the corrections can be as large as -50 GeV for $\de^{RR}_{ct}$ close to $\pm 0.5$. In the case of SPS2 they can be up to -2 GeV for this same region.

Regarding the plots involving the down-type squark sector deltas it is clear that the constraints from $B$~physics data are so tighted that the Higgs mass corrections found are very tiny.

\section{Conclusions}
\label{sec:conclusions}

We have reviewed the analysis of the 
one-loop corrections to the Higgs boson masses in the MSSM
with Non-Minimal Flavor Violation, assuming that the flavor violation is being generated from the hypothesis of general flavor mixing in the squark mass matrices. Here we have focused on the analysis of the light Higgs mass corrections that are originated from the flavor mixing between the second and third generations, and that are compatible with the constraints from  \bsg, \bmm\ and \dmbs data.

We found large corrections, mainly for the low  $\tan
\beta$ case, up to several tens of GeV for $\mh$. 
These corrections are specially relevant in the case of the light MSSM Higgs boson 
since they can be negative and up to three orders of magnitude larger than the anticipated ILC
precision of 50 MeV \cite{tesla}. Consequently, these corrections should be taken into account in
any Higgs boson analysis in the NMFV MSSM framework. 
Conversely, in the case of a Higgs boson mass measurement these
corrections might be used to set further constraints on $\deXYij$. 
The present work clearly indicates that the flavor mixing parameters $\delta^{LR}_{ct}$ and  $\delta^{RL}_{ct}$ are
severely constrained by the present bounds on the lightest Higgs boson mass within the NMFV-MSSM scenarios.

\section{Acknowledgments}

M. Arana-Catania thanks the organizers of the LCWS11 workshop for the kind invitation to present this talk and for this fruitful and enjoyable meeting at Granada.

\begin{footnotesize}


\end{footnotesize}


\end{document}